\documentclass[sn-mathphys-num]{sn-jnl}

\usepackage{graphicx}%
\usepackage{multirow}%
\usepackage{amsmath,amssymb,amsfonts}%
\usepackage{amsthm}%
\usepackage{mathrsfs}%
\usepackage[title]{appendix}%
\usepackage{xcolor}%
\usepackage{textcomp}%
\usepackage{manyfoot}%
\usepackage{booktabs}%
\usepackage{algorithm}%
\usepackage{algorithmicx}%
\usepackage{algpseudocode}%
\usepackage{listings}%
\usepackage{verbatim}
\usepackage{enumitem}
 \usepackage{comment}

\theoremstyle{thmstyleone}%
\theoremstyle{thmstyletwo}%

\theoremstyle{thmstylethree}%

\begin{document}

\title[Article Title]{Spatial Dynamics of Synchronized Ants: A Mobile Oscillator Model Analysis}


\author[1]{\fnm{Jos\'e F.} \sur{Fontanari}}

\affil[1]{\orgdiv{Instituto de F\'{\i}sica de S\~ao Carlos}, \orgname{Universidade de S\~ao Paulo},\orgaddress{ \city{S\~ao Carlos}, \postcode{13566-590}, \state{S\~ao Paulo}, \country{Brazil}}}

\author[2]{\fnm{Paulo R.A.} \sur{Campos}
}
\affil[2]{\orgdiv{Departamento de F\'{\i}sica, Centro de Ci\^encias Exatas e da Natureza}, \orgname{Universidade Federal de Pernambuco},\orgaddress{ \city{Recife}, \postcode{50740-560}, \state{Pernambuco}, \country{Brazil}}}


\abstract{Synchronized short-term activity cycles are a prominent feature in the nests of certain ant species, but their functional significance remains controversial. This study investigates whether synchronization enhances spatial accessibility within ant nests, a factor critical for efficient task performance and colony fitness. We use the mobile oscillator model, a computational framework that simulates ant movement and synchronous activity, to examine the effect of synchronization on the spatial distribution of inactive ants, which act as obstacles to the movement of active ants.  Our analysis reveals no significant effect of synchronization on the size of inactive ant clusters or on direct measures of spatial accessibility, such as the number of steps and net displacement during ant activity periods.
These findings suggest that within the mobile oscillator model, synchronization does not directly enhance spatial accessibility, highlighting the need for more complex models to elucidate the intricate relationship between synchronization and spatial dynamics in ant colonies.
}




\maketitle

\section{Introduction}\label{sec1}

Synchronized behavior is a hallmark of social insect colonies, underpinning the intricate organization and efficient functioning of these complex societies \cite{Winfree_2001,Strogatz_2004,Sumpter_2010}. Among social insects, ants exhibit a remarkable array of synchronized activities  \cite{Gordon_2019}. 
One striking example is the occurrence of synchronized  short-term   activity cycles within the nests of certain ant species, notably those belonging to the genera {\it Leptothorax} and {\it Temnothorax}. These small colonies, typically comprising a few hundred workers, exhibit rhythmic fluctuations in activity, with periods ranging from minutes to hours. 
 During these cycles, a significant portion of the colony  becomes active simultaneously, moving around inside the nest before returning to a state of relative inactivity.  Notably, ants spend approximately $72\%$  of their time in an inactive state within the nest  \cite{Franks_1990,Cole_1991,Doering_2022}.

The mechanisms driving these  short-term activity cycles involve local interactions between individuals. Physical contact plays a significant role, with active ants stimulating inactive nestmates to become active, leading to a wave-like propagation of activity through the colony \cite{Tennenbaum_2017,Anderson_2023}. The autocatalytic ant colony model provides a theoretical  framework for understanding the emergence of synchronized short-term  activity cycles, suggesting that active ants increase the activation rate of inactive ants, leading to bursts of collective movement \cite{Goss_1988}.  Individual behavioral traits, such as an ant's spontaneous activity level, its tendency to interact with nestmates and of particular relevance the resting period  also contribute to the overall synchronization patterns observed at the colony level \cite{Pedro_2024a,Pedro_2024b}.  It is  important to note that beyond these within-nest activity rhythms, ants also exhibit synchronization in other critical behaviors, where circadian rhythms and chemical communication play significant roles \cite{Das_2023}. However, this paper primarily focuses on the short-term activity cycles. 

Ant colonies of the genera {\it Leptothorax} and {\it Temnothorax} often inhabit remarkably confined spaces. Their nests can be found in pre-formed cavities such as hollow acorns, rotting sticks, or small crevices between rocks \cite{Pratt_2001,Gravish_2013}. These natural nesting sites impose significant spatial constraints on the colony members.  The challenges of navigating such confined environments are compounded by the presence of numerous nestmates, which can lead to crowding, and the fact that a substantial portion of the colony may be inactive at any given time, potentially acting as physical obstacles \cite{Doering_2023}. Consequently, spatial accessibility within these nests becomes a critical factor influencing colony efficiency and task performance.

 Spatial accessibility refers to the ease with which individuals can reach various points in their environment \cite{Kastner_2022}. Within an ant nest, this could refer to how quickly and easily an active ant can reach a particular larva, a food storage area, or any other part of the nest that requires attention, so improving spatial accessibility is likely to have positive fitness consequences for the colony. Quantifying spatial accessibility in such a complex, dynamic environment can be challenging, and one interesting solution is to measure the density of inactive ants in different sectors of the nest to assess potential obstacles to movement \cite{Doering_2023}. 

While the fitness benefits of enhanced spatial accessibility are well-established, as efficient movement optimizes essential tasks like brood care and resource distribution, the adaptive significance of synchrony remains debated \cite{Couzin_2018}. It has been suggested that synchrony may be an epiphenomenon, a byproduct of simple contact-mediated activation between ants \cite{Cole_1991}. However, if synchronized movement demonstrably improves spatial accessibility within the nest, then the observed short-term activity cycles would possess an adaptive rationale. This hypothesis, central to the work of Doering et al. \cite{Doering_2023}, provides the primary motivation for our current study.

 To examine quantitatively  the impact of synchronization on spatial accessibility, we used the mobile oscillator model \cite{Doering_2023}. In this model, each ant is represented as an oscillator with an internal clock, characterized by a phase ranging from $0^\circ$  to $359^\circ$, advancing one degree per time step and resetting to $0^\circ$  upon reaching $360^\circ$.
 The initial phase distribution dictates the instantaneous proportion of active ants, with the degree of synchrony controlled by the Kuramoto parameter, $R$ \cite{Kuramoto_1975}.  Activity and inactivity durations are determined by the parameter $\alpha$, with ants active for a fraction $\alpha/360$   and inactive for $(360-\alpha)/360$ of the time. We typically set this to $\alpha=100$, which is consistent with empirical observations that suggest about $72\%$ inactivity.  Furthermore, we introduced an asynchronous control, where activity cycle durations were drawn from an exponential distribution with mean $\alpha$. Contrary to a preliminary analysis reported in Doering et al. \cite{Doering_2023}, our findings revealed no evidence that synchronization diminishes the size of inactive ant clusters, regardless of whether cluster size was quantified via maximum local density or the largest cluster metric. Subsequently, we evaluated more direct measures of spatial accessibility, such as the number of steps and net displacement during an ant's active phase. Again, we found no significant enhancement of these metrics due to synchronization within the mobile oscillator model.

The remainder of this paper is structured as follows. 
Section \ref{sec:mod} provides a detailed exposition of the mobile oscillator model's temporal and spatial dynamics, specifically focusing on the scenario where all ants are initially inactive. Notably, Appendix \ref{appA} presents an analytical derivation of the feasible range of $\alpha$ values, ensuring consistency between initial phase assignments and a prescribed Kuramoto parameter, $R$. Section \ref{sec:so} investigates the impact of synchronization strength, $R$, on the size of inactive ant clusters, comparing results from synchronous and asynchronous simulations. Section \ref{sec:st} examines the influence of synchronization on direct measures of spatial accessibility, such as the number of steps and net displacement during an ant's active phase. Finally, Section \ref{sec:disc} synthesizes our findings and offers concluding remarks.

\section{The mobile oscillator model}\label{sec:mod}

The mobile oscillator model was introduced as a theoretical construct to test the hypothesis that the synchronous bursts of activity observed in the nests of some ant species help to improve spatial accessibility within ant nests \cite{Doering_2023}. The goal of the mobile oscillator model is not to explain the emergence of synchronized bursts of activity, as is done, for example, by the autocatalytic ant colony model introduced in the late 1980s \cite{Goss_1988} (see \cite{Pedro_2024a,Pedro_2024b} for a recent analytical studies). In fact, ant activity is synchronized by construction in the mobile oscillator mode (hence the oscillator in the model name). 

The synchronization of the ant activity is achieved by assigning each ant or oscillator $i=1, \ldots, N$ an internal clock whose phase $\theta_i$ takes an integer value in the range from $0^\circ$ to $359^\circ$. The phase advances one degree at each time step of the simulation and returns to $0^\circ$ when it reaches $360^\circ$. It is the phase $\theta_i$ that determines whether ant $i$ is active or inactive at time $t$: if $\theta_i < 360^\circ- \alpha$ then ant $i$ is inactive, otherwise ant $i$ is active. Here $\alpha$ is an integer parameter in the range $0^\circ$ to $360^\circ$, which we choose to be the same for all ants, following Doering et al. \cite{Doering_2023}. So the fraction of time the ants are active is $\alpha/360^\circ$. The transition from inactive to active and back is then a completely deterministic process. Additionally, at $t=0$ we randomly choose the phases $\theta_i$ in the range $0^\circ$ to $359^\circ- \alpha$ so that all ants are inactive at the beginning of the simulation. Of course, this scenario of total inactivity is only feasible if $\alpha < 360$ and is repeated with a period of $360$ time steps. 

From the synchronization aspect, what distinguishes one ant from another are the initial values of their phases $\theta_i$.  In the mobile oscillator model, these phases are set to produce a predetermined value of the Kuramoto  parameter
\begin{equation}\label{Kura}
R = \left |  \frac{1}{N} \sum_{j=1}^N e^{ 2 \pi \mathbf{i} \theta_j / 360^\circ } \right |,
\end{equation}
which  takes the value $1$ in the case of complete phase synchrony (i.e.,  $\theta_j  = \theta$ for all $j$)  and a value close to $0$ when the phases $\theta_j $ are chosen randomly and $N$ is very large \cite{Kuramoto_1975}.  This can be achieved by using an adaptive walk algorithm in the region of phase space where the ants are inactive (i.e., we generate random phases and keep those that bring the value of $R$ closer to the target value).  However, if $\alpha$ is large, it may be impossible to find a set of initial phases consistent with the given value of $R$. In Appendix  \ref{appA} we derive a simple expression for the maximum value  $\alpha$ can have for a given $R$. 
 In principle, $R$ measures the degree of phase synchrony between the ants, but its role in the mobile oscillator model is marginal, as we show in this paper. This is because the ants are synchronized by construction, regardless of the value of $R$, with a period of $360$ time steps. What $R$ controls is the number of ants in the activity peaks, not the synchrony between them or the average activity of the colony. 
To establish a baseline for comparison, we introduce an asynchronous control, where ant activation and inactivation are governed by probabilities $p_{inac} = 1/\alpha$  and $p_{act} = 1/(360 -\alpha)$, respectively, maintaining the same average colony activity as the synchronous model.

\begin{figure}[t] 
\centering
 \includegraphics[width=1\columnwidth]{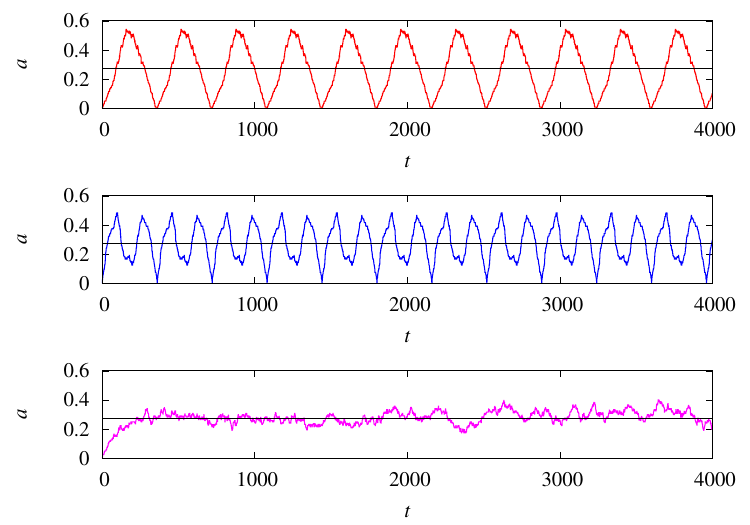}  
\caption{Proportion of active ants $a$ as a function of time $t$ for $R=0.5$ (top panel), $R=0$ (middle panel) and asynchronous dynamics (bottom panel). The thin horizontal lines show the average activity $\alpha/360 \approx 0.278$. One time unit corresponds to one degree in the internal clock of the ants, so the period of the deterministic activation dynamics is $360$ time steps. The parameters are $N=120$ and $\alpha=100$.  
 }  
\label{fig:1}  
\end{figure}

Since the spatial dynamics has no influence on the activity of the ants (i.e. on the activation dynamics), to illustrate the model features introduced so far, we show in Fig. \ref{fig:1} the temporal dynamics of colony activity  $a$ for $R=0.5$, $R=0$, and  the asynchronous dynamics. 
The colony activity $a$ is  given by the fraction of active ants at time $t$.  The initial phases for $R=0$ are generated using the adaptive walk method, since picking random initial phases typically produces larger values of the Kuramoto  parameter. We note that the asynchronous dynamics considered in Doering et al. appears to be a hybrid of our $R=0$ and asynchronous scenarios: the activity fluctuations around the mean $\alpha/360$ are  small, but the dynamics is periodic with a period of $360$ time steps, although it never returns to the initial value $a=0$ \cite{Doering_2023}.

\begin{figure}[t] 
\centering
 \includegraphics[width=1\columnwidth]{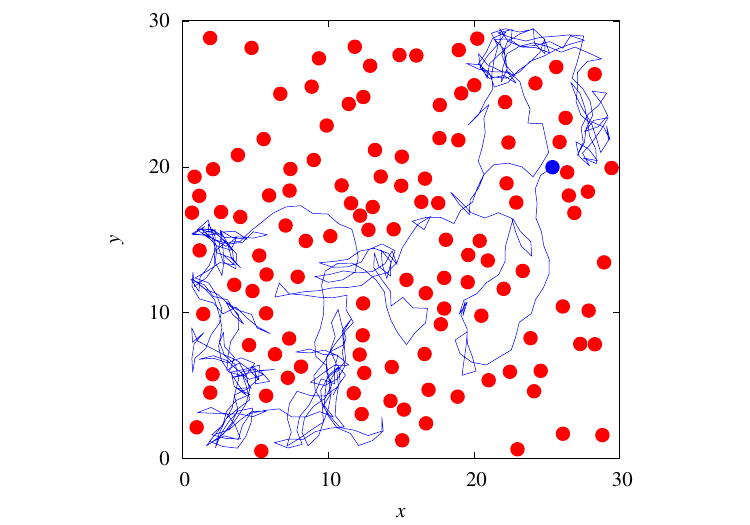}  
\caption{Illustration of the initial grid with the ants represented by unit-diameter circles. All ants except the blue ant, which is initially located at $x=25.4$ and $y=20.0$, remain fixed at their initial positions. The thin blue line is the trajectory of the blue ant for $4000$ time steps, ending at $x=10.3$ and $y=4.3$. The parameters are $\alpha=100$, $N=120$, and $L=30$.  
 }  
\label{fig:2}  
\end{figure}

We now consider the spatial dynamics of the ants. At $t=0$ the ants are randomly distributed in a square arena of linear size $L$ space units. The ants are modeled as disks with a diameter of $1$ space unit, so they cannot overlap (i.e., the distance between the centers of any two ants cannot be less than $1$ space unit), and they cannot be closer than $1/2$ space unit to the boundaries of the arena. In addition to the ants' initial positions, we assign a heading (i.e., direction of movement) to each ant by choosing randomly an angle in the range $[0,2\pi)$.  We use radians when considering the continuous angles that define directions in the plane, and degrees when considering the discrete phases of the ants' internal clocks. 

The spatial dynamics proceeds  as follows. A target ant is chosen at random.  If it is inactive, nothing happens (i.e. it stays where it is). If it is active, it chooses a direction to move in by randomly choosing an angle within $\pi/4$ to the left or right of its current heading and tries to move 1 unit distance in the chosen direction. If the new position is inaccessible, either because it is occupied by another ant or because it  is too close to the  arena boundaries, the target ant randomly chooses a new heading in the range $[0,2 \pi)$ and stays at its original position. So it misses the chance to move.  If the new position is accessible, the target ant moves there and updates its heading. Thus, each active ant performs a correlated random walk in the square arena until it encounters an obstacle. Then a new target ant is chosen and the process is repeated. The time (and the internal clocks of all ants) is updated by one unit when $N$, not necessarily different, target ants are selected to try to move. The rule that an active ant does not move when hitting an obstacle increases the blocking effect of the inactive ants, since the active ants can cluster around them, effectively increasing the size of the obstacle.

Figure \ref{fig:2} illustrates the spatial dynamics in a didactic  scenario where all ants except one, the ant initially located at $x=25.4$ and $y=20.0$, are kept fixed in their initial positions. One can see the correlated random walk performed by the single mobile ant in the regions where it can move freely and, more importantly, that it is temporarily trapped in certain regions of the arena due to obstacles created by the fixed ants and the boundaries of the arena. In fact, an indirect way to quantify the spatial inaccessibility of the ants is to measure the size of the obstacles created by the inactive ants.   

Next, we discuss two measures of obstacle size that indirectly measure spatial inaccessibility, namely the maximum local density metric and the largest cluster size, and then we consider more direct measures such as the mean number of steps taken by an ant during the activity period. Most of our results are presented for the parameter setting $N=120$, $L=30$ and $\alpha = 100$, which better matches the density and activity of real ant nests \cite{Doering_2023}.

\section{Statistics of obstacles}\label{sec:so}

Obstacles are formed by inactive ants and are easily detected and measured by taking snapshots of the nest at successive observation times \cite{Doering_2023}.  Here we consider two measures of obstacle size that provide an indirect indication of the spatial accessibility of the nest by active ants.

\begin{figure}[t] 
\centering
 \includegraphics[width=1\columnwidth]{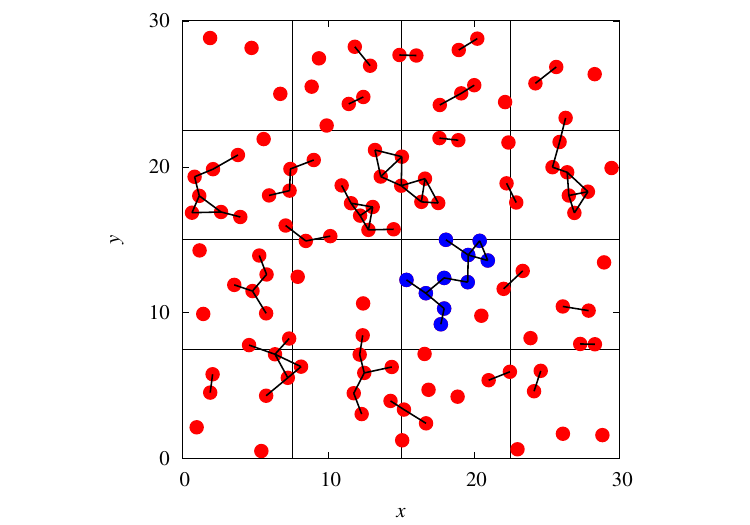}  
\caption{Clusters of inactive ants in the initial grid with the ants represented by circles of unit diameter.  Two ants are connected by a link if their distance is less than two diameters. There are 52 clusters in total, since each isolated ant counts as one cluster. The largest cluster, represented by the blue circles, contains $l=10$ ants.   The squares with side $L/4$ represent the 16 sectors used in the calculation of the maximum local density metric.  The parameters are $N=120$ and $L=30$.  
 }  
\label{fig:3}  
\end{figure}

\subsection{Maximum local density of inactive ants}

\begin{figure}[t] 
\centering
 \includegraphics[width=1\columnwidth]{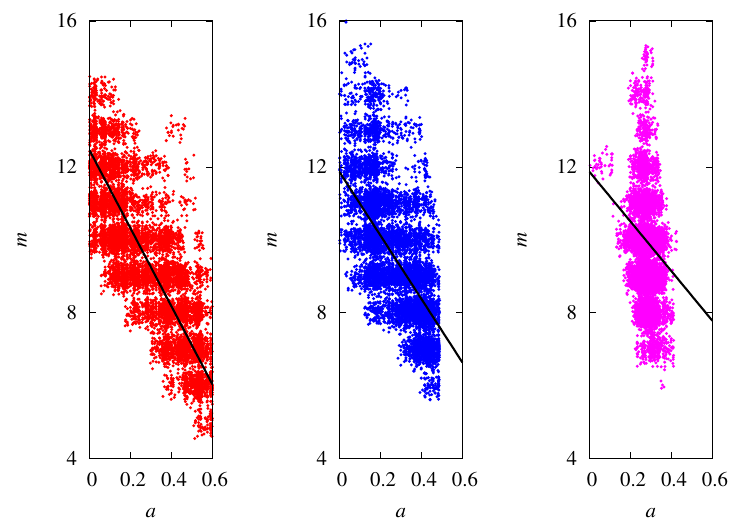}  
\caption{Scatter plots of the nest activity $a$ and  the  maximum local density  metric $m$  for the  $4000$ time step simulations of Fig. \ref{fig:1} and $R=0.5$ (left panel), $R=0$ (middle panel) and the asynchronous dynamics (right  panel). The data are  jittered for better  visualization.  The black lines in each panel are the least-squares fits of the data.  The slopes of the regression lines are $-10.7$ for $R=0.5$,  $-8.7$ for $R=0$,  and $-6.8$ for the asynchronous dynamics. The average activity of the nest is $\alpha/360 \approx 0.278$. The parameters are $N=120$, $L=30$ and $\alpha=100$.  
 }  
\label{fig:4}  
\end{figure}

In Doering et al., the maximum local density of inactive ants is used as a metric for the spatial inaccessibility of both real and simulated ant nests \cite{Doering_2023}. This metric is obtained by dividing the square arena into 16 sectors of equal size (i.e., squares of side $L/4$) and is given by the number of inactive ants in the sector with the most inactive ants (see Fig. \ref{fig:3}). Since  it is not really a density, we  denote it by the integer $m$, which  is measured at each moment of the activity time series shown in Fig. \ref{fig:1}.  Strictly speaking, the 16 sectors do not have the same area available to the ants: sectors on the edges of the grid have less free area, since the ants cannot get closer than $1/2$ space units from the edges, but we will ignore this effect.
Following Doering et al.,  
in Fig. \ref{fig:4} we show scatter plots of $a$ and $m$   from these time series. The  Pearson correlation is $r=-0.82$ for $R=0.5$, $r=-0.66$ for $R=0$,  and $r=-0.19$ for the asynchronous dynamics. Doering et al. concluded that the high negative correlation for $R=0.5$ implies a benefit to spatial
accessibility from synchrony because when more of the colony is active, there are fewer immobile obstacles throughout the entire nest \cite{Doering_2023}.  However, the main reason for these negative correlations, as well as for the large negative slopes of the regression lines, is that the number of inactive ants necessarily decreases with increasing activity, leading to a corresponding decrease in the maximum local density metric.  The difference between the correlation coefficients in the three cases shown in the figure may simply reflect the different range of activity values, especially in the case of asynchronous dynamics, in which most of the activity is restricted to the range $[0.2,0.4]$.  In fact, if we consider only the data with $a \in [0.2,0.4]$  in the left panel of Fig. \ref{fig:4}  we find that $r$ drops  from $-0.82$ to $-0.34$.  In the case of the asynchronous dynamics, we note that  the negative slope of the regression line  (see right panel of Fig. \ref{fig:4}) is not affected by the relatively few low activity data points corresponding to the beginning of the time series when all agents are inactive.

For the single runs shown in Fig. \ref{fig:4}  we find no significant differences between the means   of the maximum local density of inactive ants for the different dynamics: in all cases we find  $\bar{m} \approx 9.5$. However, the standard deviation is slightly larger for $R=0.5$ (i.e., $\sigma_m = 2.0$) compared to $R=0$ (i.e., $\sigma_m =1.8$) and the asynchronous dynamics (i.e., $\sigma_m =1.4$), which is likely a consequence of the different activity ranges of these dynamics.

The results presented so far are for a single run. To verify that synchronized activity has an effect on the maximum local density of inactive ants, we need to average over many (typically $10^4$) independent runs that differ in the initial phases of the ants' internal clocks and in the distribution of ants in the lattice at $t=0$.  The runs start with all ants in the inactive state and continue until $t=4000$. For each run we estimate the mean  $\bar{m}$ and the standard deviation  $\sigma_m$ and then we average these results over the independent runs. These averages are denoted by $\langle \bar{m} \rangle$ and $\langle \sigma_m \rangle $.

\begin{figure}[t] 
\centering
 \includegraphics[width=1\columnwidth]{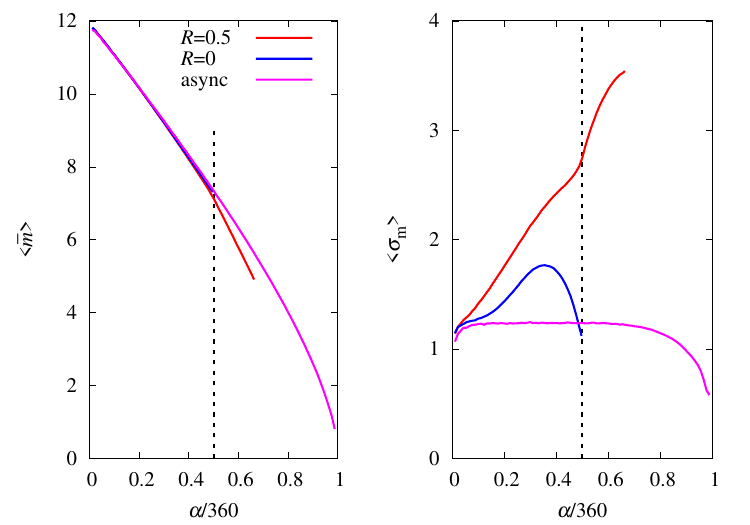}  
\caption{Mean (left panel) and standard deviation (right panel)  of  the maximum local density of inactive ants  as  function the average colony activity $\alpha/360$.  The curves for  $R=0$ and $R=0.5$ end  at $\alpha_{max} = 179$ and $\alpha_{max}= 239$, respectively. The vertical dashed lines at $\alpha = 180$ show the average activity above which the nest reaches full activity for some time intervals.  The parameters are $N=120$ and $L=30$.  
 }  
\label{fig:5}  
\end{figure}

Figure \ref{fig:5} shows the summary statistics $\langle \bar{m} \rangle$ and $\langle \sigma_m \rangle $ as a function of the fraction of time each ant is active $\alpha/360$, i.e.,  the average  activity of the colony.  Recall that the  average activity in  the single run used to generate the scatter plots of Fig. \ref{fig:4}  is $100/360 \approx 0.28$. The synchronous dynamics curves end abruptly at $\alpha =\alpha_{max}$ because there are no initial phases consistent with the fixed value of the Kuramoto parameter $R$, as discussed in the Appendix \ref{appA}.  Consistent with the observations for the single run, the summary statistics $\langle \bar{m} \rangle$ are practically indistinguishable for the three cases, except for $\alpha >180$, because the synchronous dynamics spend some time in an arena without obstacles (the nest reaches full activity for some time intervals as shown in the Appendix \ref{appA}), which reduces their mean size. In addition, the standard deviation is much larger for $R=0.5$, which is a consequence of the fact that the instantaneous activity  $a$ can reach much lower values than for the asynchronous dynamics. In fact, it even reaches the maximum $a=1$ for $\alpha >180$, causing an abrupt increase in  $\langle \sigma_m \rangle $.

\begin{figure}[t] 
\centering
 \includegraphics[width=1\columnwidth]{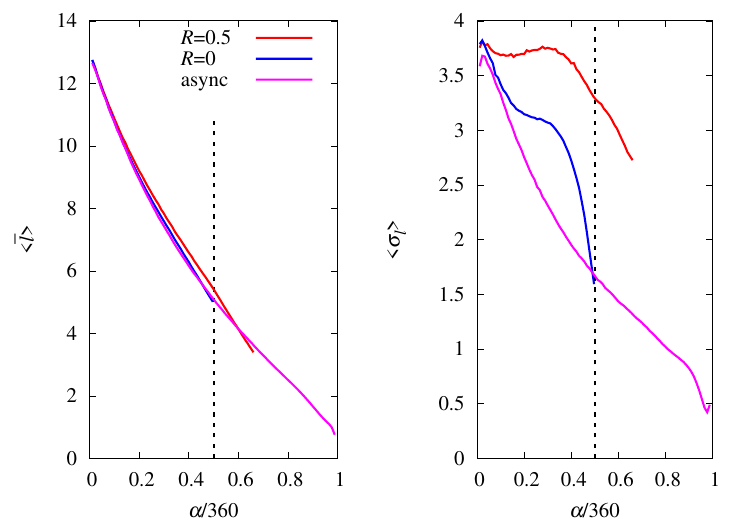}  
\caption{Mean (left panel) and standard deviation (right panel)  of  the size of the largest cluster  of inactive ants  as  function the average colony activity $\alpha/360$.  The curves for  $R=0$ and $R=0.5$ end  at $\alpha_{max} = 179$ and $\alpha_{max}= 239$, respectively. The vertical dashed lines at $\alpha = 180$ show the average activity above which the nest reaches full activity for some time intervals.  The parameters are $N=120$ and $L=30$.  
}
\label{fig:6}  
\end{figure}

\subsection{Largest cluster of inactive ants}

Since the unit-diameter discs representing the ants cannot overlap, two ants whose centers are separated by less than two diameters form an obstacle to the passage of an ant between them. This observation motivates the introduction of the concept of inactive ant clusters. These clusters are formed by connecting any two ants at a distance of less than two diameters to form an undirected graph.
The identification and labeling of clusters is efficiently done by adapting the Hoshen-Kopelman algorithm \cite{Hoshen_1976} to non-regular graphs, but our interest here is only in the largest cluster, and in particular the number of ants belonging to it, which we denote by $l$.  The advantage of this metric is that it does not depend on the particular way we divide the arena into sectors.  Figure \ref{fig:3} shows the resulting clusters for the same initial distribution of ants (all ants are inactive) used in Fig. \ref{fig:2}.     The largest cluster contains $l=10$ ants and the densest sector contains $m= 12$ ants. There are three sectors with this value, one of which happens to contain the largest cluster.

As in the analysis of the maximum local density metric, for each run we estimate the mean $\bar{l}$ and standard deviation $\sigma_l$ of the size of the largest cluster of inactive ants from $t=0$ to $t=4000$ and then average these results over $10^4$ independent runs. We denote these averages by $\langle \bar{l} \rangle$ and $\langle \sigma_l \rangle $. Figure \ref{fig:6} shows the results as a function of the  mean nest activity.  Although the effect is very small, synchronized activity creates the largest obstacles.  Similar to  the  maximum local density metric, the standard deviation is larger for $R=0.5$ than for the other dynamics. However, the dependence of $\langle \sigma_l \rangle $  on the mean nest activity is remarkably different from that shown in Fig. \ref{fig:5}, even  in the case of the asynchronous dynamics. So $m$ and $l$ are not equivalent measures of obstacle size.

Before moving on to more direct measures of spatial accessibility in the mobile oscillator model, it is appropriate to summarize our conclusions from the obstacle size analysis. Although ants that happen to be active only during bursts of activity when obstacles are smaller in principle do more work (i.e., move more)  from the perspective of potential work realization all ants are equal. Thus, the mean size of the obstacle, whether measured by $\langle \bar{m} \rangle$ or $\langle \bar{l} \rangle$, seems to be the appropriate metric to evaluate the work done by the active ants in the nest. In this sense, Figs. \ref{fig:5} and \ref{fig:6}  show that the average size of the obstacles is not significantly affected by the synchronicity of the activity.  

\begin{figure}[t] 
\centering
 \includegraphics[width=1\columnwidth]{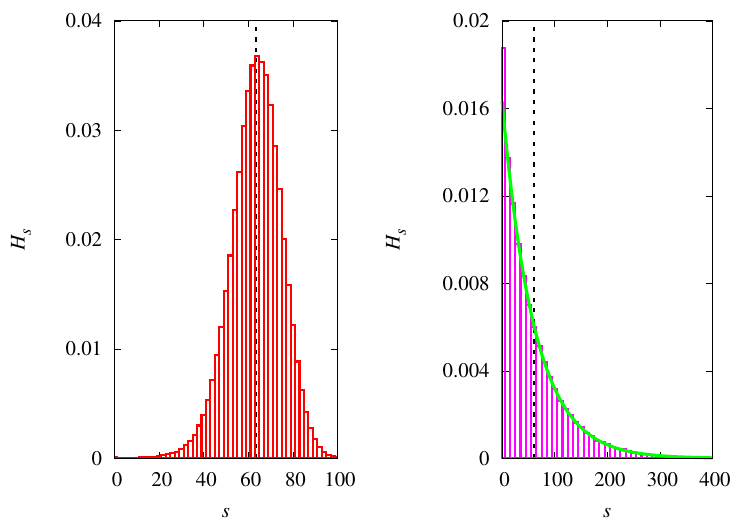}  
\caption{Distribution of steps $H_s$ taken by an ant during the activity period for $R=0.5$ (left panel) and the asynchronous dynamics (right panel).   The vertical dashed lines indicate the mean $\bar{s}$, which is two steps larger for the synchronous dynamics. The solid line in the left panel is the exponential fit. The parameters are $\alpha=100$, $N=120$ and $L=30$.  
}
\label{fig:7}  
\end{figure}

\section{Statistics of  trajectories}\label{sec:st}

To clarify whether synchronized motion can improve spatial accessibility in the  mobile oscillator model, we will use more direct measures such as the number of steps during the activity period and the distance between the ant's final and initial positions.  We are aware that while these metrics are easy to compute for the model, they are unlikely to be practical to implement in experimental studies of real ants. This practicality is the great advantage of the maximum local density metric. While it is a bit far-fetched to consider the case $R=1$ (i.e., all ants are either inactive or active) in the analysis of obstacles, because when all ants are inactive they are no obstacle to anyone, and when all are active there is no obstacle, we can consider this case in the analysis of trajectories.

If there were no obstacles (i.e., arena boundaries or other ants), the number of steps $s$ during the time an ant is active would be exactly $\alpha$ in the case of synchronous dynamics and an exponentially distributed random variable with mean $\alpha$ in the case of asynchronous dynamics. 
In the presence of obstacles, the ants take fewer steps because they miss the opportunity to move if they hit an obstacle. Figure \ref{fig:7} shows the distribution of steps $H_s$ for $R=0.5$ and the asynchronous dynamics.  This distribution was estimated by recording the number of steps taken by active ants during their activity periods in $10^4$ independent simulations, starting with inactive ants only at $t=0$ and running to $t=4000$. For the asynchronous dynamics, $H_s$ is very well fitted by an exponential distribution.  The distributions of steps for $R=0$ and $R=1$ are indistinguishable from that for $R=0.5$.

\begin{figure}[t] 
\centering
 \includegraphics[width=1\columnwidth]{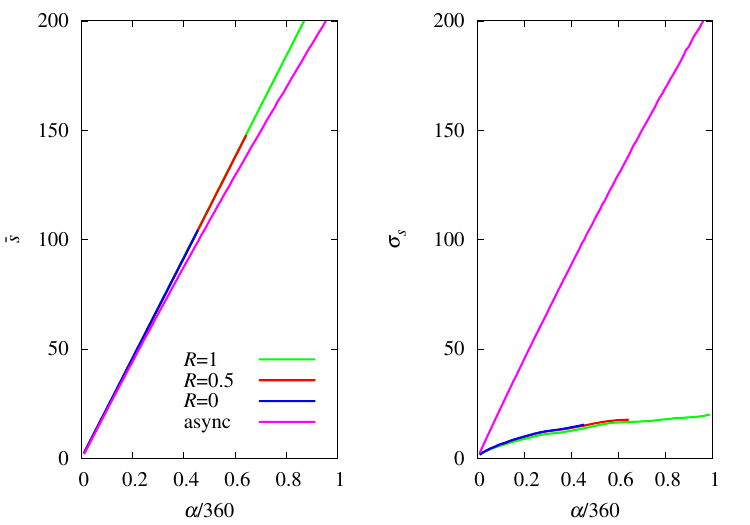}  
\caption{Mean (left panel) and standard deviation (right panel)  of the number of steps taken by an active ant  as  function the average colony activity $\alpha/360$.  The curves for  $R=0$ and $R=0.5$ end  at $\alpha_{max} = 179$ and $\alpha_{max}= 239$, respectively, and are virtually indistinguishable from the curve for $R=1$. The parameters are $N=120$ and $L=30$.  
}
\label{fig:8}  
\end{figure}

Figure \ref{fig:8} shows the mean $\bar{s}$ and the standard deviation $\sigma_s$ of the number of steps taken by an ant during the activity period as a function of the mean activity of the nest. The linear dependence of $\bar{s}$ on $\alpha/360$ is expected, since the latter gives the fraction of time an ant is active, and the more time an ant is active, the more steps it is expected to take. In particular, for the asynchronous dynamics we find $\bar{s} = 0.60 (\alpha/360)$, whereas for the synchronous dynamics we find $\bar{s} = 0.64 (\alpha/360)$.  
If there are no obstacles, we have $\bar{s} = (\alpha/360)$. This indicates that about $36\%$ of the movement attempts in the synchronous dynamics are frustrated.  The value of the Kuramoto parameter $R$ has no significant effect on $H_s$.  The mean number of steps is larger for the synchronous than for the asynchronous dynamics, especially in nests with large mean activity: the largest difference is  20 steps  for $R=1$ and $\alpha = 359$. However,  the standard deviations are so different that it makes little sense to compare the means.

\begin{figure}[t] 
\centering
 \includegraphics[width=1\columnwidth]{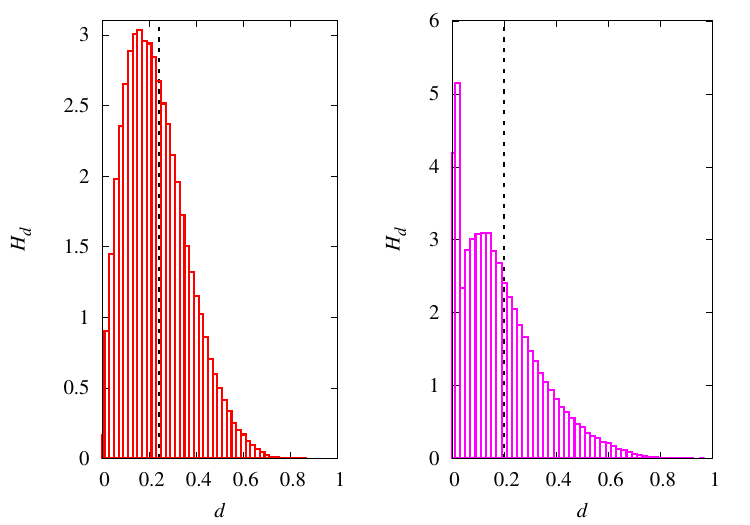}  
\caption{Distribution scaled net displacement  $H_d$  for $R=0.5$ (left panel) and the asynchronous dynamics (right panel).   The vertical dashed lines indicate the mean $\bar{d}$. The parameters are $\alpha=100$, $N=120$ and $L=30$.  
}
\label{fig:9}  
\end{figure}

A more direct measure of spatial accessibility is the distance between the position of an ant when it becomes active and the position when it becomes inactive. We define the scaled net displacement  $d \in [0,1]$ as the ratio between this distance and the largest possible distance between two ants in the square arena, i.e., $(L-1) \sqrt{2}$. Figure \ref{fig:9} shows the distribution of $d$, denoted by  $H_d$,  for $R=0.5$ and the asynchronous dynamics. The high frequency of small displacements for the asynchronous dynamics is due to the fact that there is a high probability that the ant will only be active for a very short period of time. The mean scaled net displacement is $\bar{d}=0.24$ for $R=0.5$ and $\bar{d}=0.20$ for the asynchronous dynamics, which gives a difference of only 1.6 space units in the square arena. This difference is at most 2 space units when $R$ and $\alpha$ are at their maximum values.

\begin{figure}[t] 
\centering
 \includegraphics[width=1\columnwidth]{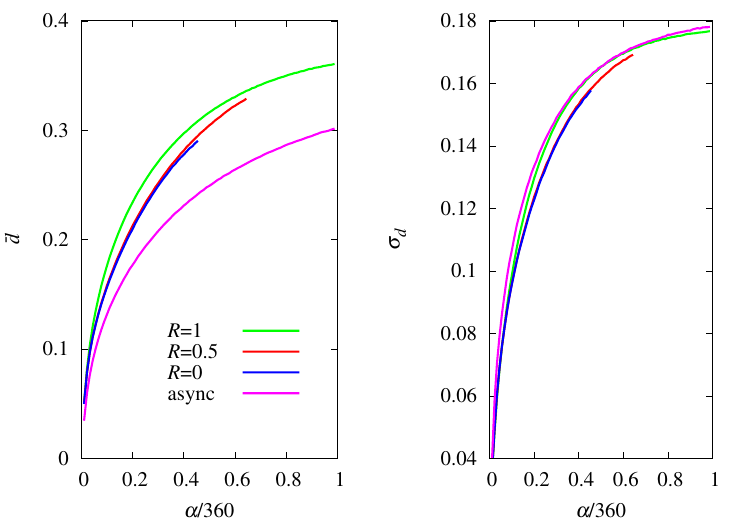}  
\caption{Mean (left panel) and standard deviation (right panel)  of the scaled net displacement of an active ant  as  function the average colony activity $\alpha/360$.  The curves for  $R=0$ and $R=0.5$ end  at $\alpha_{max} = 179$ and $\alpha_{max}= 239$, respectively. The parameters are $N=120$ and $L=30$.  
}
\label{fig:10}  
\end{figure}

Figure \ref{fig:10} shows the summary statistics $\bar{d}$ and $\sigma_d$ as a function of the mean activity of the nest. Now the standard deviations of the synchronous and asynchronous dynamics are comparable so it makes more sense to compare the means. Synchronized motion increases slightly the net displacement of the active ants: the largest difference from the asynchronous result is $\bar{d} = 0.05$ which corresponds to $\bar{d} (L-1)\sqrt{2}  \approx 2$ space units or two ant diameters. 

\begin{figure}[t] 
\centering
 \includegraphics[width=1\columnwidth]{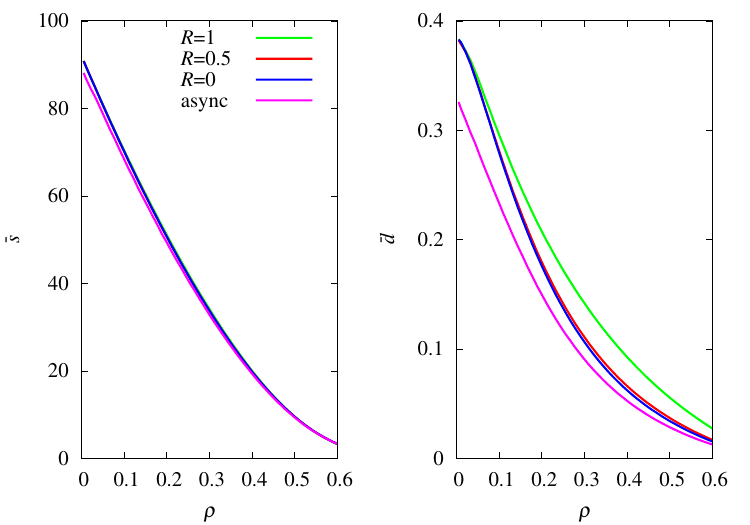}  
\caption{Mean  number of steps taken by an active ant  (left panel) and mean  scaled net displacement of an active ant (right panel)   as  function of  the density of ants in the nest $\rho=N/L^2$.   The parameters are $\alpha=100$ and $L=30$.  
}
\label{fig:11}  
\end{figure}

In addition to the average activity of the nest ($\alpha/360$), another important parameter to characterize the nest is the density $\rho= N/L^2$. 
The results presented so far  are for $\rho = 0.13$. Since the statistics of the obstacles does not allow to distinguish adequately the different dynamics, we present the effect of $\rho$ only for the statistics of the trajectories. Figure \ref{fig:11} summarizes our results. To vary $\rho$ we keep $L=30$ fixed and vary $N$.  As expected, the mean number of steps $\bar{s}$ during the activity period decreases as $\rho$ increases and is not useful to distinguish between the different dynamics. However, the mean scaled net displacement $\bar{d}$ shows some interesting features.  First, the effect of the Kuramoto parameter $R$, which determines the fraction of ants in the burst of activity as well as the duration of these bursts, is enhanced in high-density nests. Second, and more importantly, the improved spatial accessibility of synchronous dynamics has little to do with motion synchronization, as the effect is greater at low densities where synchronization is not important, as discussed next.

\begin{figure}[t] 
\centering
 \includegraphics[width=1\columnwidth]{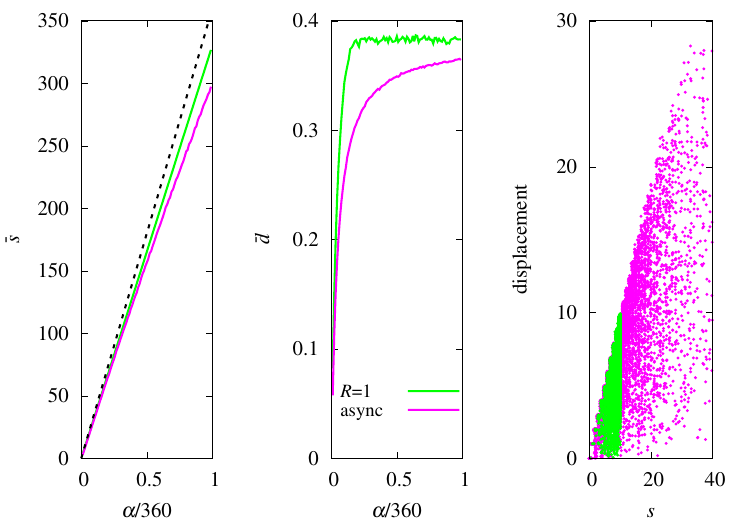}  
\caption{Mean  number of steps taken by an active ant  (left panel) and mean  scaled net displacement of an active ant (middle panel)   as a function of the proportion of time the ant is active $\alpha/360$ for $N=1$.  The dashed straight line is $\bar{s} = \alpha$. The right panel is a scatter plot of the unscaled displacement $\bar{d}(L-1)\sqrt{2}$ and the number of steps $s$ for $\alpha =10$. The data are jittered for better visualization.   The linear size of the lattice is $L=30$.  
}
\label{fig:12}  
\end{figure}

Figure \ref{fig:12}  shows the results for $N=1$, i.e., a single ant moving in the square arena,  so that synchronization has no effect at all, but still the synchronous and the asynchronous dynamics produce significantly different results. Note that in a boundless arena, the number of steps for the synchronous dynamics would be exactly $\alpha$. This scenario is described by the dashed line in the left panel. The reason that $\bar{s} < \alpha$ in the synchronous simulations  is that if the ant bumps into the walls, it will miss the opportunity to move at that time step. More importantly, the number of bumps with the walls $\alpha - \bar{s}$ increases with the length of the activity period. This observation is crucial for understanding why the mean number of steps is smaller for the asynchronous dynamics.  In this case, the  length of the activity period is an exponentially distributed random variable with mean $\alpha$. If it is less than $\alpha$, the synchronous dynamics have an advantage, and if it is greater than $\alpha$, the asynchronous dynamics have an advantage, but this advantage is reduced by the increased number of collisions with the walls that occurs for longer periods of activity. This point is illustrated in the right panel of Fig. \ref{fig:12}, which shows the scatter plot of the number of steps and the unscaled displacement for a small activity level (i.e., $\alpha=10$), so that collisions with the walls are less frequent. Overall, the net effect is a greater number of steps for the synchronous dynamics shown in Fig. \ref{fig:12}. As a result of the smaller number of steps, the mean displacement is also smaller for the asynchronous dynamics.

A word about the notation for our summary statistics is in order. In the study of obstacle statistics, we used angle brackets to emphasize the fact that we compute, for example, the mean value the maximum local density of inactive ants  $\bar{m}$ and its standard deviation $\sigma_m$ for each simulation of duration $4000$ time steps, and then average these results over $10^4$ independent simulations. So we have an average standard deviation $\langle \sigma_m \rangle$. This procedure was necessary because we introduced the  statistics $\bar{m}$ and $\sigma_m$ using the scatter plots for a single simulation. However, in the study of trajectory statistics, we calculate the mean and standard deviation by pooling data from all simulations. Of course, these different procedures only  affect the estimate of the standard deviation.

\section{Discussion}\label{sec:disc}

The functional significance of short-term activity cycles in ant nests remains a compelling enigma. A non-adaptive perspective posits these cycles as mere epiphenomena, byproducts of inter-ant interactions, potentially even detrimental to colony fitness due to interference between active individuals \cite{Cole_1991}. However, our current understanding, informed by a detailed analysis of the autocatalytic ant colony model's parameter space \cite{Pedro_2024a,Pedro_2024b}, suggests that periodic solutions are not inevitable, challenging the epiphenomenon hypothesis. Alternative adaptive explanations, such as the potential for synchronized activity to reduce redundant brood feeding \cite{Hatcher_1992}, highlight the complex interplay between synchronization and colony-level benefits.

Our study specifically investigated whether synchronization enhances spatial accessibility within the nest, a factor hypothesized to optimize worker interactions with brood and, consequently, influence colony fitness.  We used the mobile oscillator model \cite{Doering_2023}, a simplified yet insightful tool, to assess the impact of synchronization, controlled by the Kuramoto parameter $R$, on the size of inactive ant clusters, which act as obstacles to movement.  Contrary to our initial expectations, and despite accounting for increased obstacle sizes due to movement constraints, we found no significant effect of synchronization on cluster size, as measured by maximum local density (Fig. \ref{fig:5}) or largest cluster metrics (Fig. \ref{fig:6}).

Furthermore, we explored direct measures of spatial accessibility, such as the number of steps and net displacement during an ant's active phase. While the synchronous dynamics showed a marginal increase in the average number of steps, this difference was overshadowed by substantial variability (Fig. \ref{fig:8}). Similarly, net displacement exhibited only a slight advantage for synchronous activity, a difference attributable to the stochastic nature of activity cycle durations in the asynchronous control (Figs. \ref{fig:10} and \ref{fig:12}).

In conclusion, our analysis of the mobile oscillator model did not reveal a significant link between synchronization and enhanced spatial accessibility.  Although a positive result would have provided compelling proof of concept, the lack of such evidence may be attributed to the limitations of the model. We suggest that future research should focus on developing more sophisticated models that incorporate additional biological complexity. Alternatively, a group selection approach, analogous to those used to evolve efficient task allocation through response thresholds \cite{Duarte_2012,Fontanari_2024}, could be utilized to identify the dynamical regimes that optimize spatial accessibility metrics. Such investigations would provide deeper insights into the evolutionary pressures shaping synchronized behaviors and their impact on spatial efficiency within ant colonies.

\backmatter

\bmhead{Acknowledgments}

JFF is partially supported by  Conselho Nacional de Desenvolvimento Cient\'{\i}fico e Tecnol\'ogico  grant number 305620/2021-5.  PRAC was partially supported by Conselho Nacional de Desenvolvimento Cient\'{\i}fico e Tecnol\'ogico (CNPq) under Grant No. 301795/2022-3. This research is partially supported by Funda\c{c}\~ao de Amparo \`a Ci\^encia e Tecnologia do Estado de Pernambuco (FACEPE), Grant Number APQ-1129-1.05/24.

\section*{Declarations}

\subsection*{Funding} Conselho Nacional de Desenvolvimento Cient\'{\i}fico e Tecnol\'ogico and Funda\c{c}\~ao de Amparo \`a Ci\^encia e Tecnologia do Estado de Pernambuco.
\subsection*{Conflict of interest} The authors declare that they have no Conflict of interest.
\subsection*{Data availability} The manuscript has no associated data.
\subsection*{Author contribution}  All authors contributed equally to this work.

\begin{appendices}

\section{}\label{appA}

\setcounter{equation}{0}
\setcounter{figure}{0}

As mentioned in the main text, we use a simple adaptive walk algorithm to generate the initial integer phases $\theta_i = 0,1, \ldots, 359-\alpha$ with $i=1, \ldots, N$ that produce a predetermined value for the Kuramoto parameter defined by eq. (\ref{Kura}).  This range for the values of the phases guarantees that initially all ants are inactive.  Starting with a set of random phases, the algorithm proceeds as follows.  An oscillator $i$ and a new phase value are chosen at random. If the new phase brings the Kuramoto parameter closer to the predetermined target value, we update the phase of oscillator $i$. Otherwise, we keep the original phase. The procedure is repeated until the distance to the goal is less than $10^{-3}$ or the number of iterations is greater than $5000$. In the latter case, we conclude that the algorithm has not found a solution, so we generate a whole new set of initial phases and run the algorithm again.

\begin{figure}[t] 
\centering
 \includegraphics[width=1\columnwidth]{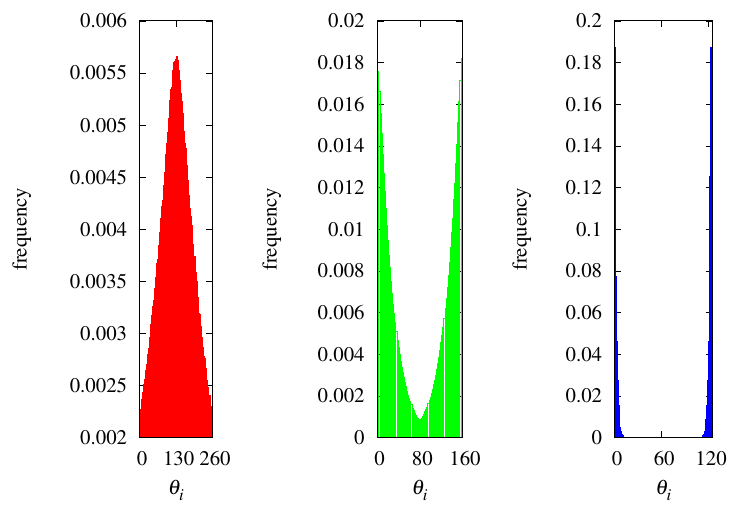}  
\caption{Distribution of  the values of the  initial  phases $\theta_i$  that solve eq. (\ref{Kura}) with $R=0.5$ for $\alpha =100$ (left panel), $\alpha =200$ (middle panel,) and $\alpha =236$ (right panel).  For $\alpha \geq 240$ there is no solution.  The number of ants is $N=120$.
 }  
\label{fig:A1}  
\end{figure}

Figure \ref{fig:A1} shows the distribution of phase values $\theta_i$ that  solve eq. (\ref{Kura}) with $R=0.5$ for three values of mean ant activity $\alpha$. It can be seen that as $\alpha$ increases, the solutions are restricted to a few values very close to the lower and upper bounds of $\theta_i$,  leading to a high degeneracy scenario where the phases take on only a few different values. In fact, for $N$ even, a solution to eq. (\ref{Kura}) with $R=0.5$  is half of the phases at the value $0$ and the other half at the value $120$. This solution exists only if $\alpha < 240$, otherwise $\theta_i$ cannot take the value $120$.   It is clear from the right panel that  the adaptive walk or any other algorithm will not find solutions to eq. (\ref{Kura}) with $R=0.5$ if $\alpha \geq 240$.

Generalizing  the above reasoning for any fixed value of $R$,  we conclude that the maximum value of $\alpha$ for which there is a set of initial phases that satisfies eq. (\ref{Kura}) is $\alpha_{max} = 359 - \theta_{max}$ where $\theta_{max}$ is given by
\begin{equation}
\cos \left (2\pi \theta_{max}/360 \right ) = 2R^2 - 1 .
\end{equation}
At this maximum half of the phases have value $0$ and half have value $\theta_{max}$.  For instance, for $R=0$ we  have $\theta_{max}=180$,   and for $R=1$ we have $\theta_{max}=0$ so the phases are homogeneous, as expected.

\begin{figure}[t] 
\centering
 \includegraphics[width=1\columnwidth]{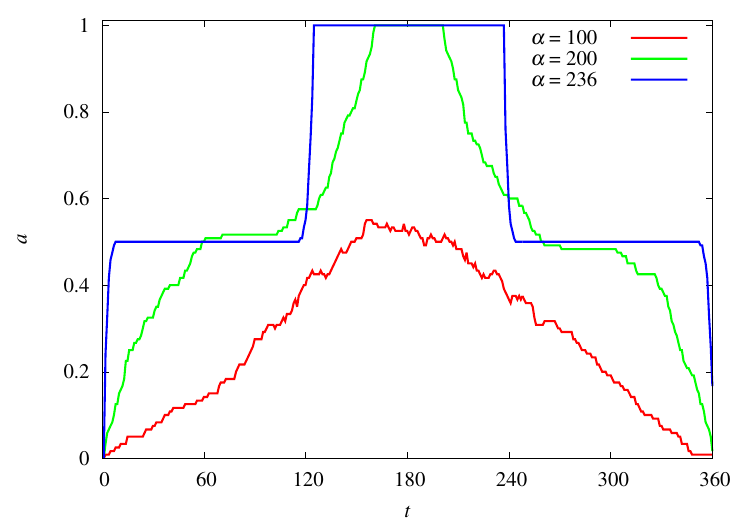}  
\caption{Proportion of active ants $a$ as a function of time $t$ for $R=0.5$ and $\alpha =100, 200$ and $236$, as indicated.  One time unit corresponds to one degree in the internal clock of the ants, so the period of the deterministic activation dynamics is $360$ time steps. The number of ants is $N=120$.
 }  
\label{fig:A2}  
\end{figure}

There is another interesting and consequential phenomenon that does not depend on the movement of the ants on the plane: there is a value of $\alpha$ beyond which all ants are active in certain time intervals, as shown in Fig. \ref{fig:A2}.  This limit is easily obtained by considering whether there is a time interval when an ant with initial phase $0$ and an ant with initial phase $359-\alpha$ are both active. For example, consider $\alpha =100$ (see left panel of Fig. \ref{fig:A1}), the ants with initial phase $0$ become active at time $t=360-\alpha = 260$ and remain active until $t=359$, while the ants with initial phase $259$ are active from time $t=1$ to $t=\alpha = 100$. Since there is no intersection between the activity periods of these two ants, the nest is never fully active. A non-empty  intersection exists provided that  $360 - \alpha \leq \alpha$, or $\alpha \geq 180$, regardless of the values of $R$ and $N$. However, for  $R=0$  the nest activity never reaches its maximum since $\alpha_{max} = 179$ in this case. 

The finding that the nest reaches full activity for $\alpha \geq 180$ is relevant to our analysis of the effect of obstacles, i.e., clusters of inactive (immobile) ants, on the spatial accessibility of the nest. Since these obstacles disappear in the full activity periods, we expect an effect on our accessibility measures based on the size of the obstacles.

\end{appendices}

\end{document}